\documentclass[aps,pra,superscriptaddress,showpacs,floatfix,12pt]{revtex4-1}
\usepackage{graphicx}
\usepackage{amsmath,amssymb}
\usepackage{amsfonts}
\usepackage{mathrsfs}
\usepackage{xcolor}
\usepackage{upgreek}
\usepackage{dcolumn}
\usepackage{amsmath,amssymb}
\usepackage{amsfonts}
\usepackage{mathrsfs}
\usepackage{epsfig}
\usepackage{dcolumn}
\usepackage{enumerate}
\usepackage{natbib}
\usepackage{amsthm}

\begin{document}

\title{Entanglement signatures for the dimerization transition in the Majumdar-Ghosh model}

\author{M. S. Ramkarthik \footnote{e-mail: ramkarthik@physics.iitm.ac.in}}
\affiliation{Department of Physics, Indian Institute of Technology Madras, Chennai, 600036, India.}
\author{V. Ravi Chandra \footnote{e-mail: ravi@niser.ac.in}}
\affiliation{School of Physical Sciences, National Institute of Science Education and Research, Institute of Physics Campus, P.O. Sainik School, Bhubaneswar, 751005, India.}
\author{Arul Lakshminarayan \footnote{e-mail: arul@physics.iitm.ac.in}}
\affiliation{Department of Physics, Indian Institute of Technology Madras, Chennai, 600036, India.}

%\preprint{IITM/PH/TH/2010/11}
\begin{abstract}
The transition from a gapless liquid to a gapped dimerized ground state that occurs in the frustrated antiferromagnetic Majumdar-Ghosh (or $J_1-J_2$ Heisenberg) 
model is revisited from the point of view of entanglement. We study the evolution of entanglement spectra, a ``projected subspace" block entropy, and concurrence in 
the Schmidt vectors through the transition. The standard tool of Schmidt decomposition along with the existence of the unique MG point where the ground states are degenerate and known exactly, suggests the projection into two orthogonal subspaces that is useful even away from this point. Of these, one is a dominant five dimensional subspace containing the complete state at the MG point and the other contributes marginally, albeit with increasing weight as the number of spins is increased. We find that the marginally contributing subspace has a minimum von Neumann entropy in the vicinity of the dimerization transition. Entanglement content between pairs of spins in the Schmidt vectors, studied via concurrence, shows that those belonging to the dominant five dimensional subspace display a clear progress towards dimerization, with the concurrence vanishing on odd/even sublattices, again in the vicinity of the dimerization, and maximizing in the even/odd sublattices at the MG point. In contrast, study of the Schmidt vectors in the marginally contributing subspace, as well as in the projection of the ground state in this space, display pair concurrence which decrease on both the sublattices as the MG point is approached. The robustness of these observations indicate their possible usefulness in the study of models that have similar transitions, and have hitherto been difficult to study using standard entanglement signatures. 
\end{abstract}
%\pacs{03.67.-a, 03.67.Bg, 03.67.Mn}

\maketitle

\newcommand{\newc}{\newcommand}
\newc{\beq}{\begin{equation}}
\newc{\eeq}{\end{equation}}
\newc{\kt}{\rangle}
\newc{\br}{\langle}
\newc{\beqa}{\begin{eqnarray}}
\newc{\eeqa}{\end{eqnarray}}
\newc{\pr}{\prime}
\newc{\longra}{\longrightarrow}
\newc{\ot}{\otimes}
\newc{\rarrow}{\rightarrow}
\newc{\h}{\hat}
\newc{\bom}{\boldmath}
\newc{\btd}{\bigtriangledown}
\newc{\al}{\alpha}
\newc{\be}{\beta}
\newc{\ld}{\lambda}
\newc{\sg}{\sigma}
\newc{\p}{\psi}
\newc{\eps}{\epsilon}
\newc{\om}{\omega}
\newc{\mb}{\mbox}
\newc{\tm}{\times}
\newc{\hu}{\hat{u}}
\newc{\hv}{\hat{v}}
\newc{\hk}{\hat{K}}
\newc{\ra}{\rightarrow}
\newc{\non}{\nonumber}
\newc{\ul}{\underline}
\newc{\hs}{\hspace}
\newc{\longla}{\longleftarrow}
\newc{\ts}{\textstyle}
\newc{\f}{\frac}
\newc{\df}{\dfrac}
\newc{\ovl}{\overline}
\newc{\bc}{\begin{center}}
\newc{\ec}{\end{center}}
\newc{\dg}{\dagger}
\newc{\prh}{\mbox{PR}_H}
\newc{\prq}{\mbox{PR}_q}
\newc{\tr}{\mbox{tr}}
\newc{\pd}{\partial}
\newc{\qv}{\vec{q}}
\newc{\pv}{\vec{p}}
\newc{\dqv}{\delta\vec{q}}
\newc{\dpv}{\delta\vec{p}}
\newc{\mbq}{\mathbf{q}}
%\newc{\dmbq}{\mathbf{\delta q}}
\newc{\mbqp}{\mathbf{q'}}
\newc{\mbpp}{\mathbf{p'}}
\newc{\mbp}{\mathbf{p}}
\newc{\mbn}{\mathbf{\nabla}}
\newc{\dmbq}{\delta \mbq}
\newc{\dmbp}{\delta \mbp}
\newc{\T}{\mathsf{T}}
\newc{\J}{\mathsf{J}}
\newc{\sfL}{\mathsf{L}}
\newc{\C}{\mathsf{C}}
\newc{\B}{\mathsf{M}}
\newc{\V}{\mathsf{V}}
\newc{\ovmg}{\overline{MG}}

%\newc{\det}{\mbox{Det}}

\section{Introduction}

Entanglement in many-body systems has been extensively studied recently \cite{amico,manybody,vidal,rvb,dagomir,latorre}
ever since the remarkable properties of quantum entanglement have come 
to be understood, especially through its various uses in quantum information processing \cite{horodecki,vincenzogate,cryptography,wootters,bennett}. 
Quantum phase transitions \cite{sachdev} which occur at zero temperature as some external parameter is changed has been particularly addressed with the help of entanglement \cite{osborne,osterloh,sun,qptent}. The Ising model critical point for example has been shown to have an entropy of entanglement that scales logarithmically  ($\sim \ln L$)  with the length, $L$, of the spin chain while away from criticality it is independent of $L$ \cite{isingcritical,holzhey,cardyentropy}. While many condensed matter systems have been studied with the help of such a ``block"  entanglement entropy \cite{vbsf,vbss,isingcritical,cardyentropy,facchi,dimerorder}, it seems more natural to consider measures of two-body entanglement like concurrence \cite{concs} in contexts where dimerization occurs \cite{secneighgu,secneichoo,heisendimer,erygit}. One of the well-studied Hamiltonians in this context is the Majumdar-Ghosh, or the $J_1-J_2$ Heisenberg model \cite{majumdarf,majumdars}. This model has a well known transition from a gapless critical phase to a gapped phase with short range correlations and dimer order \cite{okamoto,chitra}. 
Earlier studies on this model from the entanglement perspective have employed scaling behaviors of the von-Neumann entropy of contiguous blocks of spins, the valence-band entanglement entropy \cite{capponi,sun,ravindra}, and 
other measures of multipartite entanglement \cite{ujjwal} to study this transition. 

Direct signatures of the dimerization transition in this model using two-spin entanglement measures such as concurrence have been elusive. 

In the $J_1-J_2$ model the antiferromagnetic nearest neighbour Heisenberg chain is augmented with a next-nearest neighbour Heisenberg interation which is also antiferromagnetic  \cite{broekspin,majumdarf,majumdars}. 
The interest in this has been considerable since Majumdar and Ghosh proposed this as a model with an exactly solvable ground state that shows dimerization at $J_2/J_1=1/2$, the so-called MG point. The Hamiltonian is 
\beq
\label{eq:MG-Hamiltonian}
H=J_1 \sum_{i=1}^N \vec{\sigma}_i\cdot \vec{\sigma}_{i+1} + J_2 \sum_{i=1}^N \vec{\sigma}_i\cdot \vec{\sigma}_{i+2},
\eeq
with $J_1,J_2>0$. The particles are of spin-1/2 and $\vec{\sigma}_i$ are the Pauli matrices. Periodic boundary conditions are assumed, so that $\vec{\sigma}_{N+1}=\vec{\sigma}_1$ and $\vec{\sigma}_{N+2}=\vec{\sigma}_2$. At the MG point the ground state is doubly degenerate and ground state manifold is spanned by two states  $|R_N\kt$ and $|L_N \kt$, where
\beq
\label{eq:dimers}
|R_N\kt = (1\,2)(3\,4) \cdots (N-1\,N), \;\; |L_N \kt = (2\,3)(4\,5) \cdots(N\,1).
\eeq
Here, for example, $(1\,2)$ refers to the singlet state $\f{1}{\sqrt{2}}(|01\kt -|10\kt)$ of spins 1 and 2, 
and $|0\kt $ and $|1\kt$ are eigenstates of $\sigma_z$ with eigenvalues $1$ and $-1$ respectively. Thus at the MG point the degenerate ground states can be considered to have maximal nearest neighbour entanglements as the entanglement of a singlet is the maximum possible between two spin-1/2 particles. Thus it would seem natural that entanglement between spins is enhanced at a transition from a spin-liquid to a dimerized phase \cite{okamoto,chitra}, a transition which occurs when $J_2/J_1 \approx 0.24...$. While a fair amount of literature already addresses this \cite{sandvik2010,ujjwal,sun}, the present Paper revisits the issues 

from the point of view of analysis of entanglement spectra at finite lattice sizes, a non-standard block entropy, and concurrences in the Schmidt vectors of the reduced density matrices.  
 We expect this to be of interest in larger classes of problems where the possibility of transitions from spin liquid to dimer order need to be investigated.
                                             
The strategy is to focus on the fact that this happens to be one of the rare systems where at least at one point in the 
phase diagram, namely the MG point, the ground state can be solved exactly and has a form simple enough to enable
the evaluation of the entanglement spectrum analytically. The nature of the entanglement spectrum at this point
suggests a separation of the state into two orthogonal components with supports in what one may call a ``$MG$"  and
a ``non-MG" subspace. At the MG point, the $MG$ subspace {\it solely} contributes towards the construction of the ground states, hence the terminology. 
The $MG$ subspace is only 5-dimensional and in the range $0\le J_2\le 1/2$ seems dominant at least for small system sizes. 
In fact the competition between this subspace and its complement seems to be crucial for the emergence of a dimer order. 
For later convenience the non-MG subspace is denoted as $\overline{MG}$. However these subspaces are 
not unique in a way that is elaborated in the next section.
 
It is found that a suitably defined entanglement corresponding to the $\overline{MG}$ component of the wave function 
has a minimum in the vicinity of the dimerization transition. Thus while there do not seem to be simple signatures 
(except for scaling with $L$) in either the entanglement entropy of the state \cite{ravindra}, or its dominant part, namely the $MG$ component, the 
typically small $\overline{MG}$ components apparently carry information that may signal the transition. The separation of the entanglement 
spectrum into these two components also allows for a detailed study of the entanglement of the {\it eigenvectors} of the reduced density matrices. 
While much attention has concentrated on the entanglement spectrum {\it per se}, it is but natural that the eigenvectors have significant 
information in them. The entanglement in these Schmidt vectors is studied, especially the concurrence between nearest neighbors. 
It is observed that for vectors corresponding to the $MG$ component, a clear dimerization happens, 
with alternate pairs of nearest neighbour entanglements either increasing to the maximum value as the 
MG point is approached, or vanish in the vicinity of the dimerization transition.

%Thus here again simple entanglement signatures in the entanglement spectrum appear to be useful.

\section{The eigenvalue spectrum of the ground state of the MG model} 

First, a separation of the ground state of the MG Hamiltonian, say $|\Psi(J_2)\kt$, ($J_1=1$ from now) into two distinct orthogonal states,
with properties described below is sought. Thus,
\beq
\label{eq:separation}
|\Psi(J_2) \kt = \alpha(J_2) |\psi_{MG}(J_2)\kt + \beta(J_2) |\psi_{\overline{MG}} (J_2)\kt,
\eeq
where $|\psi_{MG}(J_2) \kt$ approaches a superpostion of $|R\kt$ and $|L\kt$ as $J_2 \rarrow 1/2$,
and $\alpha(J_2) \rarrow 1$, $\beta(J_2) \rarrow 0$ in the same limit. The state $|\psi_{\overline{MG}}(J_2)\kt$ is orthogonal to this, and will play a rather important role here. 
This non-MG part forms a small fraction of the whole state, at least for small $N$ (e.g. $3\%$ for $N=16$) . With increasing number of spins though this component grows and the 
detailed manner in which this happens as a function of $J_2$ is interesting and may hold information about the dimerization. However, by definition this component decreases to 
zero at the MG point of $J_2=1/2$ for {\it all} $N$. Such a separation is possible, but is potentially non-unique, as demonstrated further below.
Throughout this paper the number of spins is an even number, and there are two main subclasses: $N/2$ even and $N/2$ odd which are simply referred to as ``even" and ``odd" cases. Also from the point of view of symmetry, the translation symmetry is broken in the projected $MG$ and $\ovmg$ parts for all $J_2$.

One would especially like to treat the interval $0 \le J_2 <1/2$, 
% as the couplings remain antiferromagnetic
which contains the point where there is a gapless to gapped transition.  
When $J_2=0$ the ground state (and, indeed any, excited state) is solvable via the Bethe ansatz \cite{bethe}, 
%however the explicit forms do not seem of much use in understanding their properties. 
however the explicit forms are unwieldy and difficult to analyse in detail. 
Thus a rather ``complex"  antiferromagnetic ground state at $J_2=0$ evolves to a rather simple dimerized state at $J_2=1/2$. That a part of the ground state can be identified for all $J_2$ in the interval $[0,1/2)$
that evolves to the dimers at the MG point is not necessarily obvious and is
elucidated in this paper.%(karthik correction)

That this is possible is strongly suggested from a study of the Schmidt decomposition of the ground state. Let $N$ spins in the chain be split into two parts (say $A$ and $B$) of contiguous spins having $N_A$ and $N_B$ particles each. This paper will concentrate on the cases when $N_A$ and $N_B$ are even numbers as well. This ensures that the subsystems under consideration are of the same parity (number of spins odd/even) as the original chain. The Schmidt decomposition in terms of vectors from these two halves reads:
\beq
\label{eq:schmidt}
|\Psi(J_2)\kt = \sum_{j=1}^{2^{N_A}}\sqrt{\ld_j(J_2)} |\phi_j(J_2)\kt_A |\phi_j(J_2)\kt_B. 
\eeq
Here $\ld_j(J_2)$ are the eigenvalues of the reduced density matrix (RDM) \[ \rho_{N_A}(J_2)=\mbox{tr}_B\left(|\Psi(J_2)\kt \br \Psi(J_2)| \right),\] and $|\phi_j(J_2)\kt_A$ are the corresponding eigenvectors. The eigenvalues $\lambda_j(J_2)$ are also dependent on the partition size $N_A$, but this is not explicitly indicated. The von Neumann entropy $S_{N_A}(J_2)= -\sum_{j=1}^{2^{N_A}} \ld_j(J_2) \log(\ld_j(J_2))$ is a measure of the entanglement between parts $A$ and $B$. There have been several works that study the so-called entanglement spectrum \cite{haldane,blanc,calabrese} which is defined as 
$\{ -\ln(\ld_j), j=1,2, \ldots\}$ in many systems, such a spectrum naturally containing much more information than just the entropy. For most part of this paper, unless otherwise mentioned, $N_A=N/2$ for the even case and $N_A=N/2-1$ for the odd case. It must be noted that only for the entanglement spectrum we take logarithm to the base $e$ and for all other measures of entropy the logarithm is taken with respect to base 2. %karthik correction

\begin{figure}
%\begin{center}
%\includegraphics[height=5in,width=6.5in]{density_matrix_spectra_and_crossing.ps
%}
\epsfig{figure=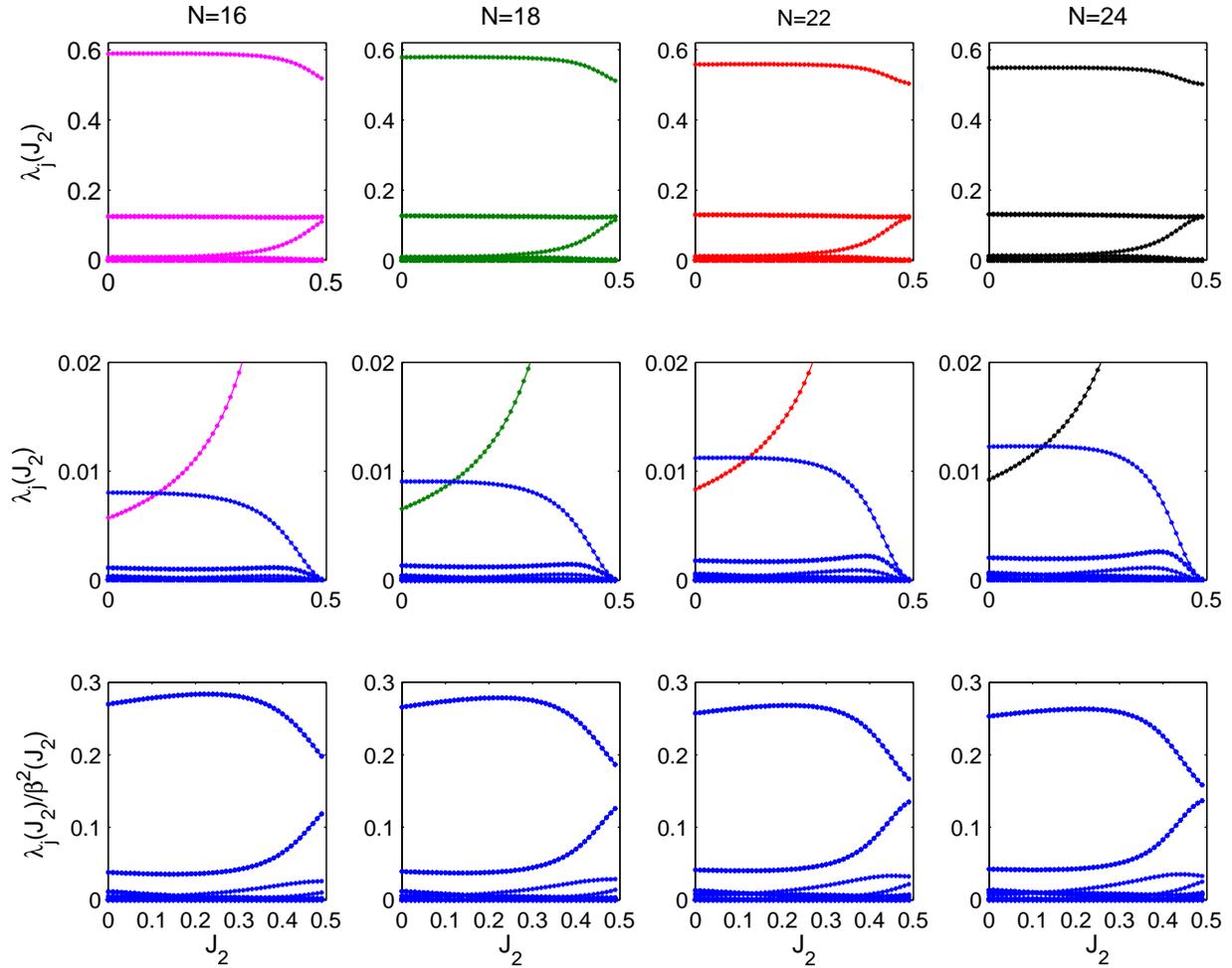, width=\linewidth}
\caption{The eigenvalues of the RDM $\rho_{N_A}(J_2)$ for $N=16$, $24$
(corresponding to $N_A=N/2$) and $N=18$, $N=22$ (corresponding to $N_A=N/2 -1$)
are shown.
Top row: 50 largest eigenvalues are plotted. Prominently seen are the  $5$
``dimer" or $MG$ subspace  eigenvalues, a large eigenvalue around $0.6$, the
almost constant triplet  around $0.1$, and the small but rising eigenvalue that
becomes important around the dimerization transition. Middle row: same as the
top, but now magnified $y$-axis, showing the $\ovmg$ triplet eigenvalue
(in one color) that crosses the rising singlet eigenvalue of the $MG$ subspace
(in a different color). Bottom row: The $\ovmg$ eigenvalues rescaled so that their sum is
unity. Largest 50 eigenvalues are shown.}
\label{entspec1}
%\end{center}
\end{figure}

Fig.~\ref{entspec1} (top row) shows the eigenvalues of the RDM, where the principal eigenvalues corresponding to the $MG$ subspace are seen clearly. The largest eigenvalue decreases as the MG point is approached from the Heisenberg.
The second largest eigenvalue actually comprises of a triplet that is almost a constant as $J_2$ varies in $[0,1/2)$.
The smallest of the eigenvalues that is clearly visible in this figure increases as the MG point is approached and indeed seems to become significant in the vicinity of the dimerization transition. It is shown below that at the MG point this eigenvalue is coupled with the largest one. As $N\rarrow \infty$, it approaches the value $1/8$. The Schmidt vectors (pure states of $N/2$ particles)  corresponding to these 5 eigenvalues along with identical vectors from the remaining $N/2$ particles form the $N-$ particle $MG$ subspace.

 Fig.~\ref{entspec1} (middle row) shows the intersection of the largest eigenvalue in the $\ovmg$ (which is triply degenerate shown using one color) with the rising eigenvalues of the dimer $MG$ sector (shown using a different color). This is a robust feature for all $N$ and an even number of spins in the subspaces. More of the eigenvalues corresponding to the $\ovmg$ subspace are seen in the bottom row which shows the rescaled eigenvalues in this sector. In the rescaled figure which shows only the $\ovmg$ eigenvalues, the most prominent ones are again few and the two that are shown correspond to a pair of triplets that seem to be coupled strongly.

\begin{figure}

\epsfig{figure=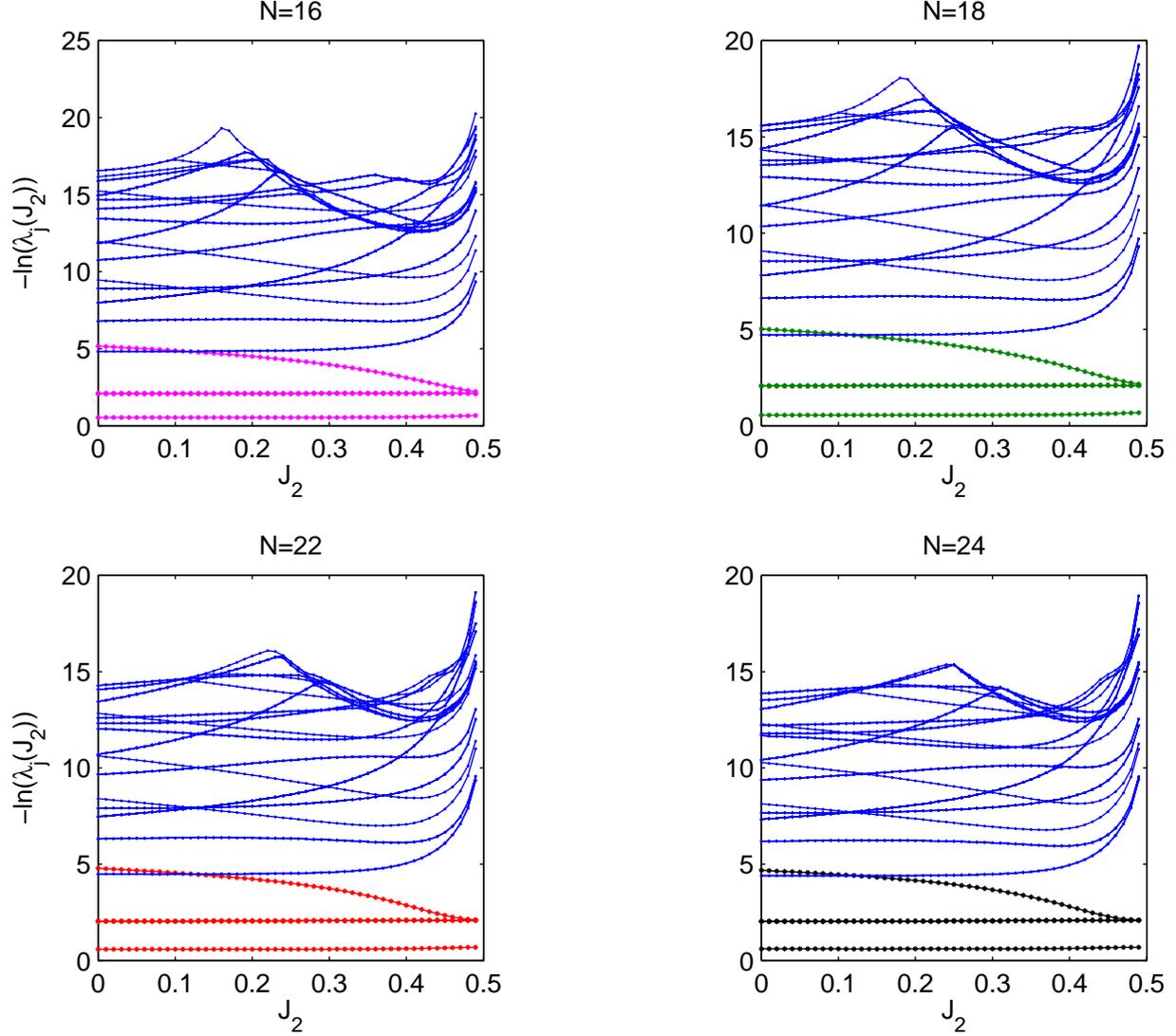, width=\linewidth}
\caption{The entanglement-spectrum that shows the $\ovmg$ as excited states (50 of the largest density matrix eigenvalues are plotted). The separation of the $MG$ and $\ovmg$ eigenvalues is seen clearly here, as well as the proliferating number of eigenvalues that vanish at isolated points along $J_2$ are seen as sharp peaks.}
\label{entspec2}

\end{figure}

  Fig.~\ref{entspec2} shows the entanglement spectrum defined as $-\ln(\lambda_j)$ plotted against $J_2$. This figure now highlights the small eigenvalues in the $\ovmg$ subspace and a clear separation is seen as those belonging to the $MG$ subspace now decrease as the MG point is approached. There are several sharp peaks that are seen in these figures and their density increases as $N$ does. 
 
These signal eigenvalues of the RDM that either go exactly to zero or come very close to it (it is sometimes difficult to tell with given numerical resolution in $J_2$) and interestingly resume their career immediately thereafter.

It is then quite apparent that there are only few dominant eigenvalues of the density matrix, even away from the MG point. That these are actually those that produce the dimer is made clear by studying the entanglement spectrum of a superposition of the dimers $|R\kt$ and $|L\kt$. 
Towards this end consider the state
\beq
\label{eq:dimersuperpose}
|\Psi_{MG}\kt = \alpha_1 |R_N\kt + \alpha_2|L_N\kt
\eeq
where $|R_N\kt$ and $|L_N\kt$ are as defined in Eq.~(\ref{eq:dimers}), and
$\alpha_{1,2}$ are real. It is the simplest type of a ``Valence Bond State",
which is a superposition of dimerized states \cite{beach}. While in general such
VBS states have been quite extensively studied, including from the point of view
of entanglement \cite{vbsf,vbss,vbskore,katsura,vbsmamb} to
our knowledge a detailed analysis of the simple state in $|\Psi_{MG}\kt $ at finite $N$ and arbitrary partition sizes has not been reported. 

We begin here by evaluating the required RDMs. 
Let $N_A=2k$ be the number of particles in the subsystem $A$ ($k$ is any appropriate integer $>1$) whose
density matrix is given by (details are relegated to an Appendix A):
\beq
\begin{split}
\rho_A^{MG}& =\mbox{tr}_B\left( |\Psi_{MG} \kt \br \Psi_{MG}|\right)=\alpha_1^2 \vert R_{2k} \kt \br R_{2k} \vert+\alpha_2^2 \bigg[ \dfrac{I_{1}}{2} \otimes |L_{2k-2}\kt\br L_{2k-2}| \otimes \dfrac{I_{2k}}{2} \bigg]  \\
& + \dfrac{\alpha_1 \alpha_2}{2^{(N-2k)/2}}(-1)^{(N-2k)/2}\bigg[\vert R_{2k} \kt \br L_{2k}| + \vert L_{2k} \kt \br R_{2k}|\bigg],
\end{split}
\label{dimerRDM}
\eeq
where $|R_{2k}\kt =(1\,2)\cdots(2k-1\, 2k)$ and $|L_{2k-2}\kt=(2\,3)\cdots(2k-2\,2k-1)$ while $|L_{2k}\kt = |L_{2k-2}\kt (2k\, 1)$ are dimers
of part $A$; $|L_{2k-2}\kt$ does not contain the singlet between the first and the ``last" ($2k$) spin of part $A$. 
As the inner product $\br R_{K}|L_K\kt = (-1)^{K/2}/2^{K/2-1}$, it is readily verified that $\mbox{tr}(\rho^{MG}_A) =\al_1^2+\al_2^2+2 \al_1\al_2 \br R_N |L_N \kt = \br \Psi_{MG} |\Psi_{MG} \kt$. Thus if $\al_1$ and $\al_2$ are taken such that $|\Psi_{MG}\kt$ is normalized the trace
of the RDM $\rho^{MG}_A$ is indeed $1$.  

To find the spectrum of $\rho^{MG}_A$, it is useful to express the identity operator in the space of spins $1$ and $2k$, $I_1 \otimes I_{2k}$  in
terms of the complete set of corresponding Bell state projectors. This results in:
\beq
\begin{split}
\label{eq:dimerRDMBell}
\rho^{MG}_A \, =\, &\al_1^2 |R_{2k}\kt \br R_{2k}| +\df{\al_2^2}{4}  |L_{2k}\kt \br L_{2k}|+\df{\al_2^2}{4} \left(\sum_{l=1}^3 |L_{2k}^l\kt \br L_{2k}^l | \right) +\\& \dfrac{\alpha_1 \alpha_2}{2^{(N-2k)/2}}(-1)^{(N-2k)/2}\left( | R_{2k} \kt \br L_{2k}| + | L_{2k} \kt \br R_{2k}| \right).
\end{split}
\eeq
Here $|L^l_{2k}\kt$ are $|L_{2k-2}\kt \otimes |\phi_b^l\kt_{2k,1}$, where
$|\phi_b^1\kt =(|01\kt+|10\kt)/\sqrt{2}$, $|\phi_b^2\kt =(|00\kt+|11\kt)/\sqrt{2}$, and $|\phi_b^3\kt =(|00\kt-|11\kt)/\sqrt{2}$ are three Bell states, the remaining one being the singlet that along with $|L_{2k-2}\kt$ results in $|L_{2k}\kt$. These Bell-state augmented dimers are quite easily seen to satisfy the following: $\br R_{2k}|L^l_{2k}\kt =\br L_{2k}|L^l_{2k}\kt = 0$, for $l=1,2,3$.
Thus $|L_{2k}^l\kt $ are three degenerate eigentates of $\rho^{MG}_A$ with eigenvalues $\ld^{MG}_2=\ld^{MG}_3=\ld^{MG}_4= \alpha_2^2/4$.
Two other eigenstates are linear combinations of the nonorthogonal states $|R_{2k}\kt $ and $|L_{2k}\kt$,
and the resultant eigenvalues are 

\beq
\label{dimerevals12}
\ld^{MG}_{1,5} = \df{1}{2} \left[ \left(\br \Psi_{MG} |\Psi_{MG}\kt -\df{3 \al_2^2}{4}\right) \pm \sqrt{ \left(\br \Psi_{MG} |\Psi_{MG}\kt -\df{3 \al_2^2}{4}\right)^2 -\left(1-\f{4}{2^{2k}}\right)\left(1-\f{4}{2^{N-2k}}\right) \al_1^2 \al_2^2} \right].
\eeq

These $5$ eigenvalues of the $2k-$particle ($k>1$) RDM are the only nonzero ones, and it is easily verified that  they add up to the trace of the RDM. If the initial dimer state is normalized they add to unity. They are ordered according to their typical magnitude, especially when 
$\al_1^2=\al_2^2=1/2$, with $\ld_1^{MG}$ being the largest and $\ld_5^{MG}$ being the smallest eigenvalue.
When $N=\infty$, and $\al_1^2=\al_2^2=1/2$, the eigenvalues are, as indicated earlier,  $(1/2,1/8,1/8,1/8,1/8)$ and the entropy or entanglement is $2 \, \mbox{ebits}$. For finite $N$ the entropy is smaller, for example when $N=8$ and $2k=4$, taking the normalized state $|\Psi_{MG}\kt$  with $\al_1=\al_2=2/3$ leads to $\ld_2^{MG}=\ld_3^{MG}=\ld_4^{MG}=1/9$ as
the eigenvalues for the degenerate triple of states and the other two are $\ld^{MG}_1=(2+\sqrt{3})/6 \approx 0.622$
and $\ld^{MG}_5=(2-\sqrt{3})/6 \approx 0.044$, while the entanglement is $\approx 1.683\, \mbox{ebits}$. 
To take an example of an odd case, let $N=10$ and $2k=4$, the normalized state that has momentum $\pi$, maybe taken  as $|\psi_{MG}\kt = \sqrt{8/17} ( |R_{10}\kt -|L_{10}\kt)$. Thus with $\alpha_1=-\alpha_2=\sqrt{8/17}$, the above gives $\ld^{MG}_{2,3,4}= 2/17\approx 0.117$ and $\ld^{MG}_1=(11+2 \sqrt{19})/34 \approx 0.579$ and $\ld^{MG}_5= (11-2 \sqrt{19})/34 \approx 
0.067$. This may be compared with the eigenvalues of the RDM near the MG point. For example for $N=10$ and $2k=4$ when $J_2=0.4975$, the triplets in the dimerized part have eigenvalues of $0.1175$, while the large eigenvalue and the smallest one in this part are $0.582$ and $0.065$, which indeed compare well with the numbers derived above.

Thus the ``entanglement-spectrum" at the MG point consists of only $5$ levels.
The fact that the ground state of the MG model for $0\le J_2<1/2$ does not
undergo any crossings \cite{sun} indicates a certain robustness that will be
reflected in the entanglement-spectrum as well. As evidenced also by results
shown in Figs. 1 and 2,  indeed a 5-dimensional subspace dominates the
entanglement spectrum and evolves to the one derived above when $J_2\rarrow
1/2$. Thus Eq.~(\ref{eq:schmidt}) maybe split into two parts with $|\Psi(J_2)\kt
= \sum_{j=1}^{5}\sqrt{\ld_j(J_2)} |\phi_j(J_2)\kt_A |\phi_j(J_2)\kt_B  +
\sum_{j=6}^{2^{N_A}}\sqrt{\ld_j(J_2)} |\phi_j(J_2)\kt_A |\phi_j(J_2)\kt_B$,
which is the separation that is alluded to in Eq.~(\ref{eq:separation}). Thus
$|\psi_{MG}J_2)\kt$ is in the 5-dimensional ``dimer"  $MG$ subspace that
dominates the state, while $|\psi_{\ovmg}(J_2)\kt$ belongs to the
$(2^{N_A}-5)$-dimensional subspace which constitutes the rest. The eigenvalues
$\ld_i(J_2)$ for $1\le i \le 5$ are {\it defined} as those that evolve to
$\ld^{MG}_i$ at the MG point $J_2=1/2$. Thus it follows that
\beq
\label{eq:Schmidt2}
\begin{split}
\al^2(J_2)&=\sum_{j=1}^5 \ld_j(J_2),\;\; \beta^2(J_2)=\sum_{j=6}^{2^{N_A}}\ld_j(J_2)=1-\al^2(J_2),\\ |\psi_{MG}(J_2)\kt =&
\sum_{j=1}^{5}\sqrt{\f{\ld_j(J_2)}{\al^2(J_2)}} |\phi_j(J_2)\kt_A |\phi_j(J_2)\kt_B, |\psi_{\overline{MG}}(J_2)\kt=
 \sum_{j=6}^{2^{N_A}}\sqrt{\f{\ld_j(J_2)}{\beta^2(J_2)}} |\phi_j(J_2)\kt_A |\phi_j(J_2)\kt_B.
 \end{split}
\eeq
The identification of the eigenvalues belonging to $\ovmg$ is complicated slightly by the fact that the largest eigenvalue in this set crosses the eigenvalue that becomes $\ld_5^{MG}$ of Eq.~(\ref{dimerevals12}). Indeed this ``rising" eigenvalue in the dimer subspace is coupled to the largest eigenvalue state and its dominance in the spectrum seems correlated with the dimerization process. 

It is important to note that the identification of the $N$-particle pure states $|\psi_{MG}(J_2)\kt$ and $|\psi_{\ovmg}(J_2)\kt$ from the Schmidt vectors 
is {\it dependent} on the partition sizes $N_A$ and $N_B$ and thus usage of terms like $MG$ and $\ovmg$ subspaces, is predicated upon a definite partition dependence, usually the symmetric one, corresponding to $N_A=N_B$.

\begin{figure}
\includegraphics[scale=0.62]{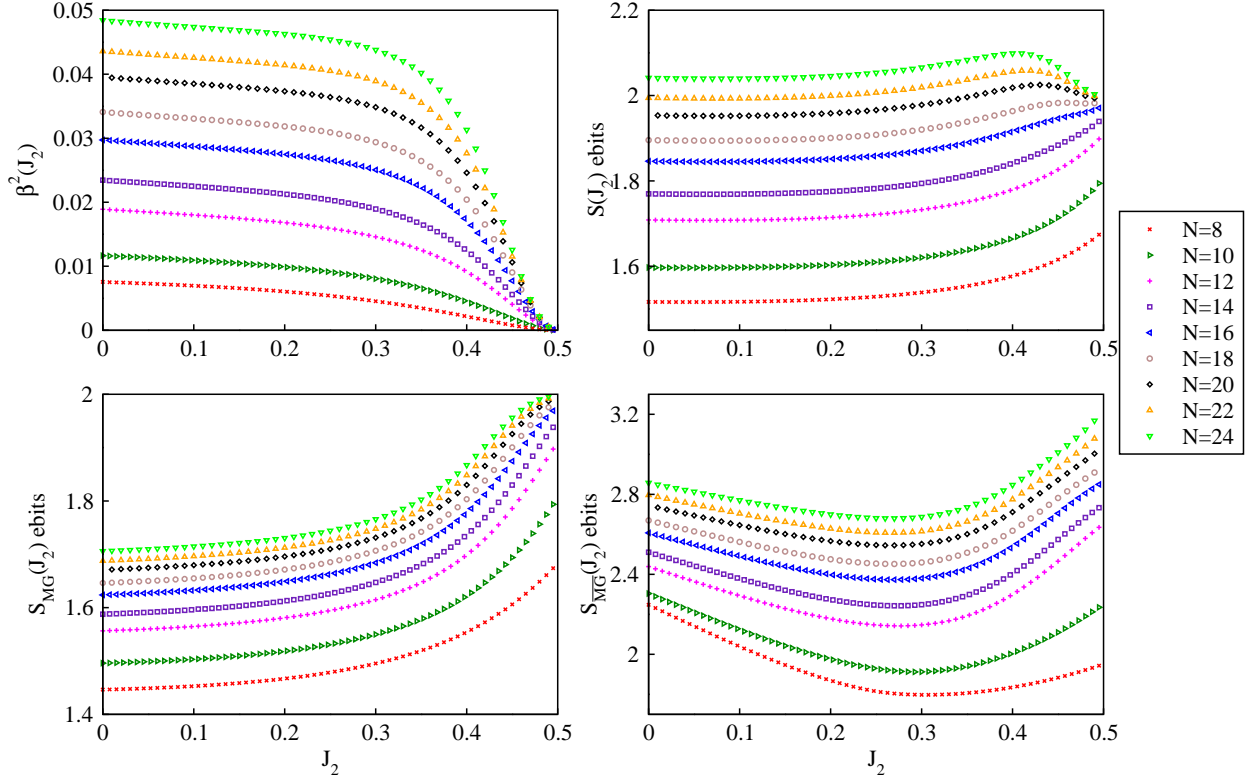} 
\caption{The sum of the eigenvalues corrsponding to the $\ovmg$ subspace, $\beta^2(J_2)$, is shown on the top left plot for various values of the number of spins $N$ from $N=8$ to $N=24$ in steps of two, while the top right plot shows the  entanglement of $N_A$ spins with the rest (the entropy $S(J_2)$) in the complete ground state . The bottom left plot shows the entanglement of the projection in the $MG$ subspace (the entropy $S_{MG}(J_2)$). The entropy $S_{\ovmg}(J_2)$, the entanglement of the projection in the $\ovmg$ subspace as defined in Eq.~(\ref{grass}) is shown in the bottom right plot. $N_A=N/2$ for the even case and $N/2-1$ for the odd.} %new 
\label{entropyplot}
\end{figure}

Following the above considerations, one may find three entanglements between $N_A$ contiguous spins and the rest: 
\beq
\label{grass}
\begin{split}
S(J_2)&=-\sum_{i=1}^{2^{N_A}} \ld_i(J_2) \log(\ld_i(J_2)),\;
S_{MG}(J_2)= -\sum_{i=1}^{5} \f{\ld_i(J_2)}{\al^2(J_2)} \log\left(\f{\ld_i(J_2)}{\al^2(J_2)}\right),\\
S_{\overline{MG}}(J_2)&= -\sum_{i=6}^{2^{N_A}} \f{\ld_i(J_2)}{\beta^2(J_2)} \log\left(\f{\ld_i(J_2)}{\beta^2(J_2)}\right),
\end{split}
\eeq
whose interpretations respectively as the entanglements in the ground state ($|\Psi(J_2)\kt$), and separately in the $MG$ and $\ovmg$ parts ($|\psi_{MG}(J_2)\kt$ and $|\psi_{\overline{MG}}(J_2) \kt$) of the ground state is straightforward. The behaviors of $\beta^2(J_2)$ and $S(J_2)$ are shown in the Figs.~(\ref{entropyplot}) top left and right plots respectively,
while the bottom left plot of Fig.~(\ref{entropyplot}) shows $S_{MG}(J_2)$, the entanglement in the $MG$ subspace projection. The bottom right plot shows $S_{\overline{MG}}(J_2)$. In these figures for even cases, ($N/2$ is even), $N_A=N/2$, while in the odd case $N_A=N/2-1$. It is interesting that while the entropies, $S(J_2)$ and $S_{MG}$ are monotonic, the entropy $S_{\overline{MG}}$
shows a minimum in the vicinity of the dimerization transition. 
If indeed these are entanglement signatures of this quantum phase transition, it is interesting that it is found in the ``non-MG" part of the state. Of course this part increases in dominance as $N$ increases, see Fig.~(\ref{entropyplot}) (top left plot). 

That there is a fairly significant dimerized part that is already present in the small $N$ Heisenberg model maybe the reason why the reason why the entanglement signatures of the transition are not easy to see; but once the dimerized part is excised, at least in part, the remaining ``grass" seems to reveal the transition.  It should also be noted that calculations not presented show that if the entropy $S$ is itself split into  a $MG$ and $\ovmg$ part without rescaling the eigenvalues, then these are monotonic on $[0,1/2]$; the interpretation of $S_{\overline{MG}}$ as a entanglement is necessary.

\begin{figure}
\flushleft
\begin{center}
\includegraphics[scale=0.52]{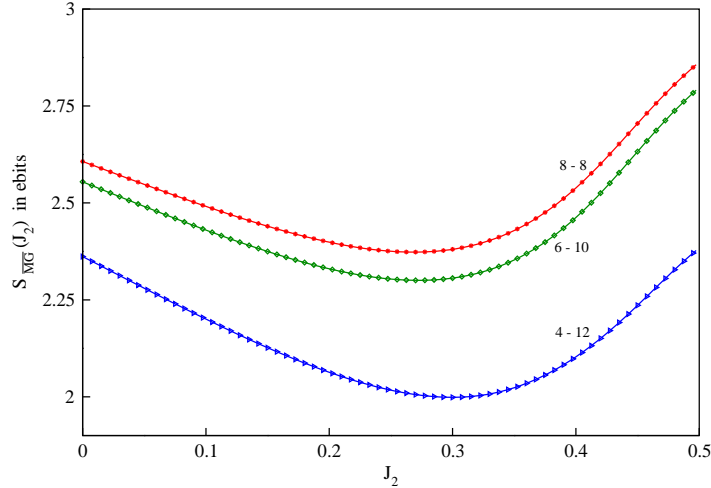}
\caption{The entropy $S_{\ovmg}(J_2)$ for $N=16$ spins with different bipartition sizes $N_A-N_B$ that are indicated. This is then the entanglement of $N_A$ spins with the rest for the $\ovmg$ projection. The existence of a minimum is robust to altering paritition sizes.}
\label{fig:entropysplit}
\end{center}
\end{figure}

It is also observed that the spectrum of the RDM for various partition sizes $N_A=2k$ are qualitatively similar including the crossing of the lowest eigenvalue corresponding to the $MG$ subspace with the triplet from the $\ovmg$ one. The existence of a minimum entropy of entanglement for the $|\psi_{\ovmg}(J_2)\kt$  state between $N_A$ and the rest of the spins is interestingly a robust feature, as shown in Fig.~(\ref{fig:entropysplit}). 
Of course for $k=1$ (two-spin RDM) is special, the number of eigenvalues of the RDM being $4$ and all remain non-zero at the MG point. In fact it is easy to see from the Eq.~(\ref{dimerRDM}), which becomes a Werner state, that these eigenvalues correspond to the one dominant one and the triplet. The ``rising"  state is absent from this spectrum and is a property of chains with more than $4$ spins. It is quite essential that the number of spins in the subsystems $A$ and $B$ are even. If there are an odd number of spins (in the subsystems) the number of eigenvalues in the RDM that are non-zero at the MG point is $4$ and the entropy $S_{\ovmg}$ remains monotonic in $[0,1/2]$. There is also the added complication that the ground state of the $J_1-J_2$ model in this range has zero momentum when $N/2$ is even and momentum $\pi$ otherwise. 

The eigenvalues in $\ovmg$ themselves have structure and a hierarchy that is not unlike that of the dimerized state.  
While the largest triplet $\ovmg$ eigenvalue, $\lambda_6(J_2)$ decreases monotonically in $[0,1/2]$, the scaled value (divided by $\beta^2(J_2)$, see the bottom panel of Fig.~(\ref{entspec1})) shows a single peak again in the vicinity of the dimerization transition, which maybe the origin of the minimum in the entropy of the grass. Indeed $-\log(\lambda_6(J_2)/\beta^2(J_2))$ is the so-called min-entropy, $S_{\infty \, \ovmg}$, and along with the von-Neumann entropy is a special case of the Renyi entropies. It is guaranteed from general considerations that $S_{\infty\, \ovmg}<S_{\ovmg}$.

\begin{figure}
\begin{center}
\includegraphics[scale=0.53]{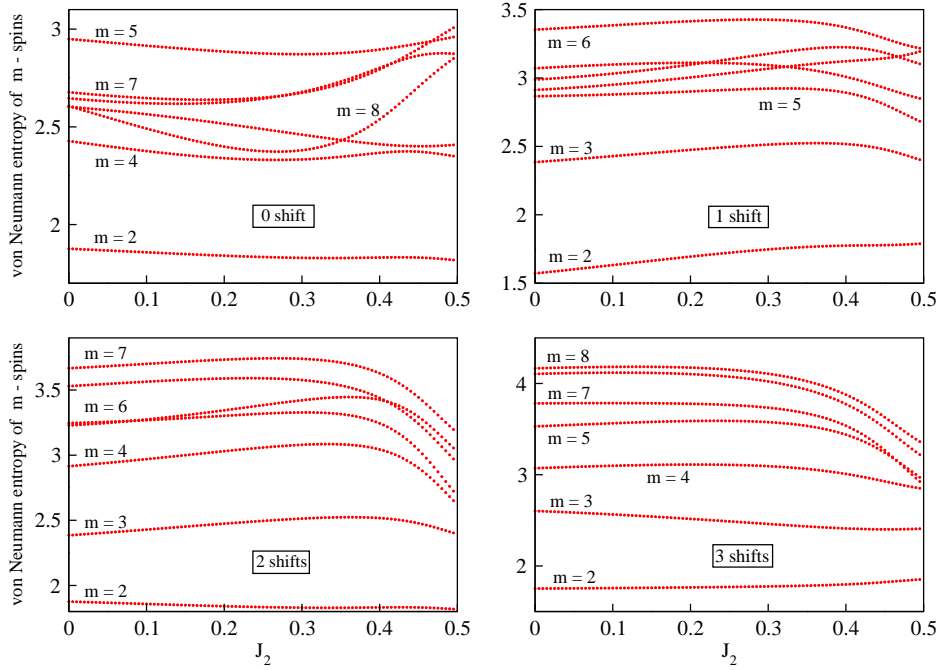}
\caption{The entanglement entropy  between $m$ contiguous spins and the rest for the state $|\psi_{\ovmg}\kt$. Here $N=16$, and the ``shift" refers to the first spin site in the block of $m$ spins. }
\label{kbits}
\end{center}
\end{figure}

To explore entanglement sharing in the pure $N-$ particle states $|\psi_{\ovmg}\kt$ and  $|\psi_{MG}\kt$,  for a fixed partition $N=N_A+N_B$, one can study multipartite measures and 2-spin measures such as concurrence. The latter is studied in the next section, while for the first, the entanglement entropy of $m$ contiguous spins is numerically calculated, in the case of symmetric partitions $N_A=N_B$, the results of which are shown in Fig.~(\ref{kbits}) for a chain of length $N=16$. The translation symmetry is lost on projection and by construction, from the Schmidt decomposition, these states have a symmetry of shifting by $N/2$ sites. Thus it makes a difference as to where the first of the $m$ contiguous spins is chosen. The case of ``0 shift"
corresponds to the first being also the first in the block of $N/2$ spins that remains after tracing. Further shifts refer to right shifting the first spin in the block by the indicated amount.  
The rather more complex entanglement sharing of the state $|\psi_{\ovmg}\kt$ is seen here. The dependence on the  shift of the first spin of the $m$ blocks is clear and for no shift the prominent feature remains the $m=8$ case that has already been discussed above. However the case of $m=4$ also shows a local minimum, albeit a shallow one, in the entanglement at exactly the same value of $J_2$ as for $m=8$. The other values of $m$ also indicate the fair amount of multipartite entanglement present in this state.
Shifting the first spin away, now explores different entanglement features, for instance with a shift of $1$, and $m=2$, this is the entanglement of spins $2$ and $3$ with the rest when the whole state is $|\psi_{\ovmg}\kt$. The first observation is that there is no minimum anymore for any value of $m$, especially $8$ spins. Thus, it is required for the minimum in the entropy that the block coincides with the partitioning in the Schmidt decomposition. The second is the considerably large entanglements that is present, for example with blocks shifted by $3$. These are in sharp contrast to the behavior of the corresponding quantities for the state  $|\psi_{MG}\kt$, wherein the entanglements are seen to be monotonically increasing and the shifts do 
not change the features much, and the entanglement entropies are only about half as large.

\section{Concurrences in the Schmidt vectors and the $MG$, $\ovmg$ projections.}

Attention is now turned to a more detailed study of two-spin entanglements. In particular one wishes to know the nature of entanglement in the eigenstates of the RDM of $\rho_{N_A}$. These correspond to the $N_A$-particle Schmidt vectors in a Schmidt decomposition of the ground state. Also of interest is the concurrences present in the corresponding projected $N-$ particle pure states $|\psi_{MG}\kt$ and $|\psi_{\ovmg}\kt$. The concurrence is a one-to-one function of the entanglement of formation and the 
recipe to obtain the concurrence between any pair of spins which are either in a
pure or a mixed state $\rho$ is as follows \cite{concs}: compute the
eigenvalues of the matrix $\rho (\sigma^y \otimes \sigma^y)
\rho^*(\sigma^y \otimes \sigma^y)$ (the complex conjugation being done in
the computational basis). The eigenvalues are guaranteed to be positive and if
they are  arranged as $\{\lambda_1 \ge \lambda_2 \ge \lambda_3 \ge \lambda_4\}$,
the concurrence between the pair of spins considered is given by
$C=\mbox{max}[0,\sqrt{\lambda_1}-\sqrt{\lambda_2}-\sqrt{\lambda_3}-\sqrt{
\lambda_4}]$. The concurrence $C$ is such that $0 \leq C \leq 1$, with
zero for the case of an unentangled state and unity when it is maximally
entangled.

Recall that the $MG$ subspace for a given even partition is spanned by $5$ states whose corresponding eigenvalues are a large and decreasing one, three degenerate and nearly constant ones, while the last is small and increasing. Concentrating on the case $N_A=N/2$ and $N$ an even integer, it is sufficient to study the $N/2-$ particle pure states  $|\phi_1(J_2)\kt_A$, $|\phi_{2,3,4}(J_2)\kt_A$, and $|\phi_5(J_2)\kt_A$ (see Eq.~(\ref{eq:Schmidt2})) respectively. Collectively they contribute to the normalized state $|\psi_{MG}(J_2)\kt$. The complementary subspace is the normalized state $|\psi_{\overline{MG}}(J_2)\kt$ whose
principal contribution comes from a triplet whose eigenvalue is decreasing and intersects with the increasing lowest eigenvalue from the $MG$ subspace. In all of these states one can look at the nature of pairwise entanglement via nearest neighbor pairwise concurrence \cite{concs}, which is a genuine and well-used measure of entanglement between two qubits or spin-1/2 particles, especially useful when they are not in a pure state. One may study the concurrences $C_{(i,i+1)}(J_2)$ between spin at $i$ and $i+1$ (identifying $L+1$ as the first spin), as well as their totals either over the entire chain, or over two parts, where $i$ is even or when it is odd. Note that while the ground state has translational invariance, this is typically broken in the states $|\psi_{MG}(J_2)\kt$ and $|\psi_{\ovmg}(J_2)\kt$, and the corresponding Schmidt vectors. Thus $C_{(i,i+1)}(J_2)$ are typically different for different values of $i$, unlike in the original state.

\begin{figure}
\begin{center}
\flushleft
\includegraphics[scale=0.61]{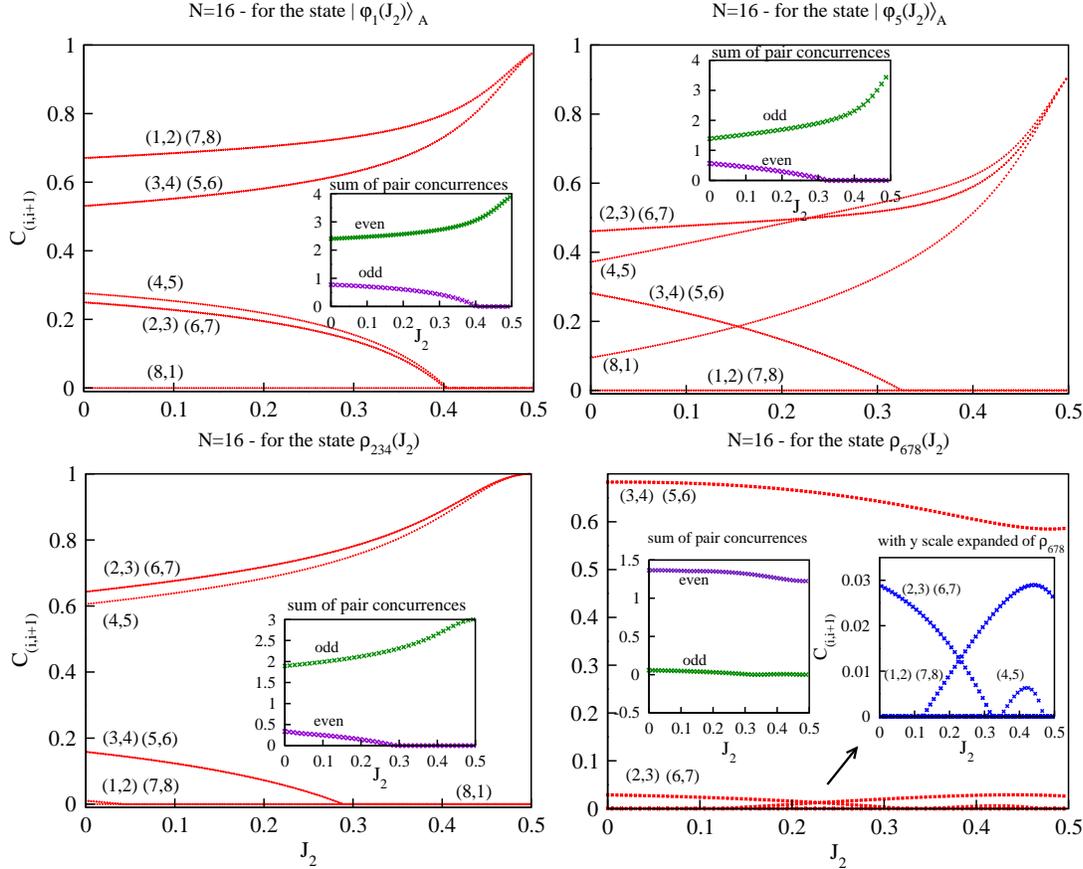}
\caption{The concurrence between neighboring spins in the Schmidt vectors, which are pure states of $N/2$ spins, for the symmetric partition case of $N=16$. The top left  is for the vector $|\phi_1(J_2)\kt$ corresponding to the highest eigenvalue,  top right is for the vector $|\phi_5(J_2)\kt$ corresponding to the rising eigenvalue, while the bottom left is for the mixed state $\rho_{234}$ corresponding to the triplet. The bottom right plot is for the mixed state $\rho_{678}$ corresponding to the largest triplet of eigenvalues corresponding to states in the $\ovmg$ subspace. The insets in the corresponding graphs show the sum of the alternate pair concurrences for each of the above described eigenstates.}
\label{conc} 
\end{center}
\end{figure}

As the dimer part of the state survives till the MG point, it is  likely to have large pairwise concurrences. The top panel of Fig.~(\ref{conc}) shows these for the most dominant state, namely $|\phi_1(J_2)\kt_A$
and the rising state $|\phi_5(J_2)\kt_A$. It is seen that the nearest neighbor concurrences show a clear progress to dimerization as $J_2$ increases. In the case of $|\phi_1(J_2)\kt_A$, the entanglement between the alternate bonds starting from the first is large and increases with $J_2$, while the others decrease and vanish well before the MG point, and in this respect is like $|R_{N/2}\kt$; while for $|\phi_5(J_2)\kt_A$ the highly entangled bonds start from the second spin, and in this respect is like $|L_{N/2}\kt$. The insets show the sum of the concurrences in the even and odd sublattices of the $N/2$ spin chain, and it is seen that the entanglements vanish in the alternate
bonds again in the vicinity of the dimerization transition, but not at exactly one point. Also notice that for $|\phi_5(J_2)\kt_A$ entanglement develops for distant spins at sites $1$ and $8$ for the case
shown of $N=16$. The full chain has been ``cut" keeping sites 1-8 and tracing out 9-16. This singles out the sites 1 and 8; also note the reflection symmetry that is apparent from the distribution of concurrences amongst the spins 1-8.  
It maybe noted that the concurrences for the case of the ground state do not show such structures that reveal the dimerization \cite{secneichoo,heisendimer,sun} 

In case of the triplets (as in the $MG$ subspace, or the states corresponding to the largest eigenvalue in the $\ovmg$ subspace) the states are not unique due to degeneracy. However the projector onto the degenerate three-dimensional subspace is unique and can be used to define a density matrix, for example for the triplet in the $MG$ subspace consider the state: 
\beq 
\rho_{234}(J_2)=\f{1}{3}(|\phi_2(J_2)\kt_{A}  ~_A \br \phi_2(J_2)|+|\phi_3(J_2)\kt_A ~_A\br \phi_3(J_2)|+|\phi_4(J_2)\kt_A ~_A \br \phi_4(J_2)|),
\eeq and the corresponding mixed state for the most prominent triple of $\ovmg$, say $\rho_{678}$. The concurrence in the bonds of these states are shown in the lower panel of Fig.~(\ref{conc}). The state $\rho_{234}$ displays large entanglements in the alternative bonds starting from the second spin and is in this respect like the rising state $|\phi_5(J_2)\kt$, except that there is here no entanglement between the distant spins 1 and 8. This distinction from the rising state is understood on calculating  the two-qubit reduced density matrix of the triplet density matrix at $J_2=1/2$:
\beq
(\rho_{234}(J_2=1/2))_{18}=\dfrac{1}{3}\left( |00\kt \br 00|+|11\kt \br 11| + |++\kt \br ++| \right)
\eeq
the reduced density matrix of the spins $1$ and $8$ is $(\rho_{234}(J_2=1/2))_{18}$, the state $|+\kt =(|0\kt +|1\kt)/\sqrt{2}$. The separability of these spins in this density matrix is then evident, and entanglement  appears to be absent not only at $J_2=1/2$ but for the entire range considered. 
The entanglement of other bonds decrease from the Heisenberg point and vanish again well before the MG point. Indeed the point where the dimer states takes on a pure alternate bond entanglement is again in the region of the dimerization transition. 

A similar analysis for $\rho_{678}$ is shown in the same figure and presents a somewhat different picture, with the dimerization not being uniformly present. While the concurrence between 
$3-4$ and the symmetric $5-6$ spins are large, the rest of the nearest neighbor entanglements are nearly zero. The concurrences that is present in the $3-4$ pair is also decreasing from the Heisenberg chain as the MG point is approached. It is observed that the entanglement between the $1-2$ and $7-8$ pairs, which starts at zero, develops as $J_2$ increases and is non-zero at $J_2=1/2$. This indicates the existence of some dimerization in the $\ovmg$ subspace, but of a different kind than in the $MG$ subspace. The effective overall decrease of the two-qubit entanglements is in sharp contrast to that found for the Schmidt vectors that span the $MG$ subspace.

\begin{figure}
\begin{center}
% \flushleft \hspace{3cm}
\includegraphics[scale=0.7]{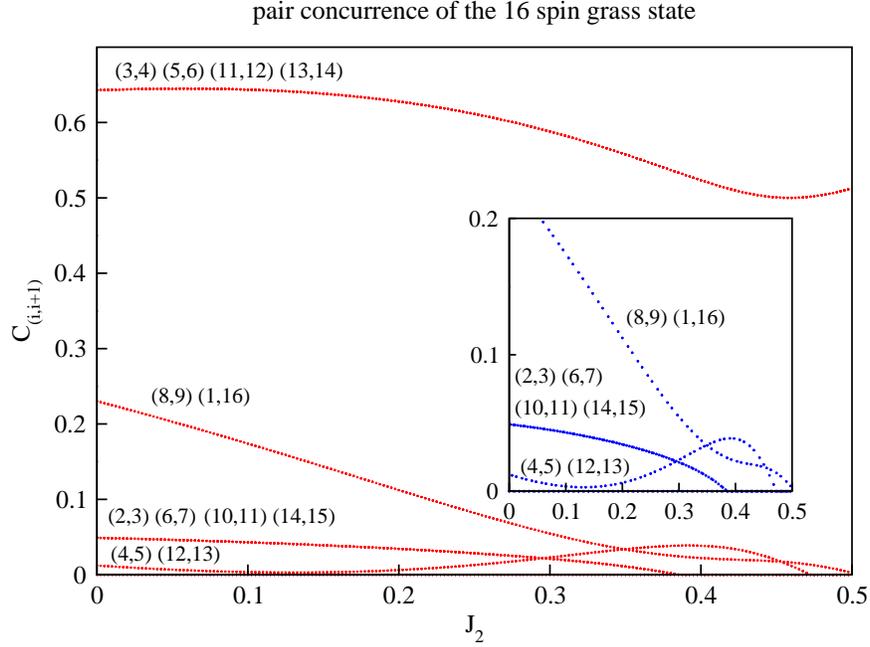}
\caption{The graph shows the nearest neighbor concurrence in the $N=16$ spin state $\psi_{\overline{MG}}(J_2)$. The spin pairs are indicated above their corresponding curves and the inset is a magnification. The pairs not indicated, namely $(1,2), (7,8), (9,10),(15,16)$, have zero concurrence throughout.}
\label{nspinconc}
\end{center}
\end{figure}

To analyze this further, pair concurrences were calculated for the $N$ spin state $|\psi_{\ovmg}(J_2)\kt $ and the nearest neighbor pair concurrence is shown in Fig.~(\ref{nspinconc}). While this looks similar to the case of the state $\rho_{678}$,
which is indeed the dominant part of $|\psi_{\ovmg}(J_2)\kt $, the prominent difference is the somewhat large entanglement between the $1-16$ and $8-9$ pairs (here $N=16$) which exists for the Heisenberg chain and which decreases away as the MG point is approached. Once again the overall decrease in the concurrence is in contrast to that for the $MG$ subspace and is consistent with an increase of the von Neumann entropy in as much as one can think of monogamy of entanglement being operative and the entanglement becomes of a more multipartite kind. Indeed the structure of even the two-spin entanglements present in this state is neither of a $|L_N\kt$ kind nor of a $|R_N\kt$ kind, but rather a mixture with some bonds being either very weakly or not at all entangled. For example the entanglement present in $1-16$ and $8-9$ is consistent with a $|L_N\kt$ kind of dimerization, while the 
prominent entanglement between $3-4$, and $5-6$ resembles $|R_N\kt$. It is interesting that as the dimerization progresses, even in this ``grass" 
contribution there is a tendency to choose a type of dimerization, with the $|L_N\kt$ kind taking a backseat at around the dimerization transition. However also the decrease in the concurrence is consistent with a rising entropy as observed in Fig.~(\ref{entropyplot}) (bottom right plot). The principal features discussed above have been verified to remain intact for the case of $N=24$ spins.

\section{Discussions and Summary}

In this paper the frustrated $J_1-J_2$ antiferromagnetic Majumdar-Ghosh model has been revisited with a view on entanglement properties, both multipartite and those between pairs of spins. Entanglement studies of the ground state that have revealed signatures of the dimerization transition have hitherto relied on scaling of the entropy with the system size. However in this paper several suggestive simple signatures are presented, from those that involve von-Neumann entropy to concurrence between spins. For this the principal tool is the well-known Schmidt decomposition that combined with the existence of the unique MG point ($J_2/J_1=1/2$) provided an opportunity for a projection of the ground state into two orthogonal subspaces that are unique once the partition in the Schmidt decomposition is fixed. The dominant subspace is only $5$ dimensional and contains the complete state at the MG point. The complementary subspace whose significance wanes from the Heisenberg point ($J_2/J_1=0$) contributes only marginally, but this contribution increases with $N$, the number of spins; for instance for $N=16$ this contribution is roughly $3\%$ when $J_2/J_1=0$. These subspaces are indicated as $MG$ and $\ovmg$, although again the partition dependence in the Schmidt decomposition is  implicit.

The entanglement between $N_A$ spins and the rest of the spin chain is known to have different scaling laws as criticality is lost at $J_2/J_1 \approx 0.24$. What is shown above is that while the entanglement 
of the full or dominant projection in the $MG$ subspace is monotonic, the projection onto $\ovmg$ 
has a minimum in the vicinity of the transition for even $N_A$, at least for the values of $N$ that has been explored. This feature is robust against various different partitions of the ground state. How robust this feature is to increasing number of spins remains to be seen. 

\begin{figure}
%\flushleft \hspace{2cm}
\begin{center}
\includegraphics[scale=0.5]{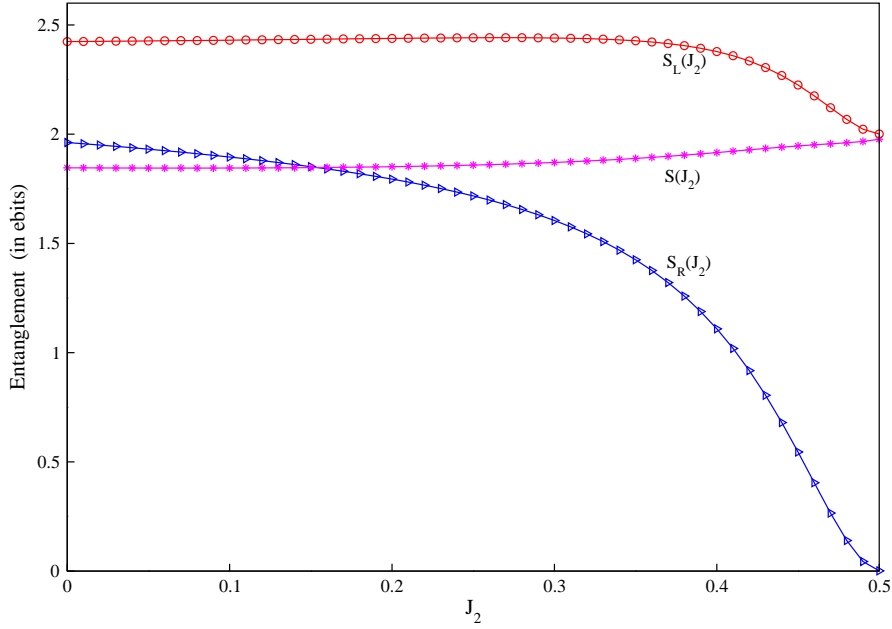}
\caption{The etanglements of half the chain with the rest for  $S_R(J_2)$ and $S_L(J_2)$ after projecting out the fully dimerized $|L_N\kt$ or $|R_N\kt$ states respectively. Shown also is the case for the complete ground state entanglement $S(J_2)$ for comparison, and here $N=16$ spins are used.}
\label{projectedout}
\end{center}
\end{figure}

To understand better the behavior of entropies, one may calculate further the entanglements present after the states $|L_N\kt$ or $|R_N\kt$ are projected out from the ground state $|\Psi(J_2)\kt$. That is the quantities $S_L(J_2)=S((|\Psi(J_2)\kt -|R_N\kt \br R_N|\Psi(J_2)\kt )t)$ and $S_R(J_2)=S((|\Psi(J_2)\kt -|L_N\kt \br L_N|\Psi(J_2)\kt )t)$, the entanglements of $N/2$ contiguous spins with the rest ($t$ is a normalization constant), are found and plotted in Fig.~(\ref{projectedout}). Note that $S_L(J_2)$ tends to the entanglement of the state $|L_N\kt$ at the MG point, namely $2$, while $S_R(J_2)$ tends to that of $|R_N\kt$, namely $0$. It is interesting to note that the entanglement of these symmetry broken states are now monotonically decreasing already for small values of $N$, reflecting well the fact that the entanglement sharing in the spins is changing from a more complex situation at the Heisenberg point to the fully dimerized one at the MG point. The dimerization, which is leading to the formation of couples that are unentangled with any other spin discourages multipartite entanglement.

An important complementary view of entanglement sharing is provided by calculating the
concurrence between pairs of spins. The state in which these are measured are however not the ground state itself, but the eigenfunctions of the reduced density matrix, or the Schmidt vectors. 
The Schmidt vectors (now pure states of $N_A$ spins) of the $MG$ subspace
show a clear progress towards dimerization as $J_2$ increases with the
the most dominant state resembling $| R_{N/2} \kt$ and the rising state, 
$| L_{N/2} \kt$. The triplet also shows dimerization as in $|L_{N/2}\kt$ except for the end spins being unentangled. 
Here ``dimerization" is seen as the vanishing of concurrence on a sublattice, while the complementary one develops 
into pairs with maximum concurrence.
The most dominant eigenvalue corresponding to triply degenerate states in the $\ovmg$ subspace was also studied using pair
concurrences and it presents a different picture compared to the $5$ Schmidt vectors in the $MG$ subspace, in that the concurrences tend to decrease as the MG point is approached. The projection of the state on the $N$ particle $\ovmg$ subspace, $|\psi_{\ovmg}(J_2)\kt$ also shows interesting differences and larger multipartite entanglements.
The initial decrease of the entropy $S_{\ovmg}$ which contributes to the non-monotonic character of this entropy may have its origins in the overall tendency for decreasing entropy 
as evidenced on projecting out the $|L_N\kt$ or $|R_N\kt$ states, however more study is warranted on the exact origins and significance, if any, of this.

If a bipartite split with one block containing the spins at odd sites and the other block containing the spins at even sites (``comb entanglement") is taken, it presents a complex entanglement spectrum with many crossings, and while this is interesting, the dimerization transition seems difficult to unravel. Also the present study has calculated non-nearest neighbor concurrence in the various states presented, but most of them 
are indeed zero.
Preliminary investigations of the $J_1-J_2$ model with quenched disorder in $J_2$ reveals
a certain robustness of the above analysis. Small disorders lead to the exact crossing at the 
$MG$ point being replaced by an avoided one, and there is still to a large extent 
only a $5$ dimensional dominant space and hence a split into an $MG$ and $\ovmg$ subspaces
persists. Further study is needed on how non-dimerized subspaces such as $\ovmg$  dominate in the large $N$ limit.

\section{Acknowledgements}
Early discussions with R. Shankar are gratefully acknowledged.
MSR thanks UGC - CSIR India for fellowship. AL and MSR acknowledge generous assistance through the 
DST-Project SR/S2/HEP-12/2009. 
\appendix
\section{Computation of the reduced density matrix and its spectrum at the MG point} 

There is more than one way to derive the reduced density matrix, and in the following a direct 
approach is used.
First start with the superposition $|\psi\kt=\alpha_1|R_N\kt+\alpha_2|L_N\kt$ of
the two dimer states that are eigenfunctions at the MG point, namely $|R_N\kt$
and $|L_N\kt$ (as given in Eq.~(\ref{eq:dimers})).

On taking a bipartite split of $2k$ contiguous spins, the block $A$, and the
remaining $N-2k$ spins, the  block $B$. One traces over the block $B$ to find
the reduced density matrix:
\begin{equation}
\rho_{2k}=\mbox{tr}_{B}(|\psi\kt\br\psi|) =\alpha_1^2 \rho_{2k_1} + \alpha_2^2 \rho_{2k_2}+
\alpha_1\alpha_2( \rho_{2k_3}+\rho_{2k_3}^{\dagger}),
\end{equation}
where $\rho_{2k_1} =\mbox{tr}_{B}(| R_N\kt\br R_N|)$, $\rho_{2k_2} =\mbox{tr}_{B}(|L_N\kt\br L_N|)$, and $\rho_{2k_3} =\mbox{tr}_{B}(|R_N\kt\br L_N|)$.

Denote now $|R_{2k}\kt \equiv (1 2) (3 4) \cdots (2k-1~ 2k)$ and $| L_{2k-2}\kt  \equiv (2 3) (4 5) \cdots (2k-2~ 2k-1)$; these being spin states which are not affected by the partial tracing operation.
It is straightforward to calculate $\rho_{2k_{1}}$ as no singlet ``bonds" are cut due to the structure of $|R_N \kt$. However for calculating $\rho_{2k_{2}}$ the singlets between $(N 1)$ and $(2k~ 2k+1)$ are broken which results in maximally mixed states $\dfrac{I_1}{2}$ and $\dfrac{I_{2k}}{2}$ at the ends. The remaining tensor products of singlets $|L_{2k-2}\kt$ are left unaffected. Thus it follows that
\beq
\rho_{2k_{1}}=\alpha_1^2 |R_{2k} \kt \br R_{2k}|, \;
\rho_{2k_{2}}=\dfrac{\alpha_2^2}{4} \bigg(I_1 \otimes |L_{2k-2}\kt \br L_{2k-2}| \otimes I_{2k}\bigg).
\eeq
The remaining part involves cross terms, which is written explicitly by introducing standard $\sigma_z$ basis 
for the spins in block $B$:
\begin{equation}
\rho_{2k_{3}}= \sum_{i_{2k+1},\cdots,i_N \in \{0,1\}} \br i_{2k+1} \cdots i_N|R_N\kt \br L_N| i_{2k+1} \cdots i_N \kt
\label{rhocross}
\end{equation}
It is easy to verify that $\br i_1 i_2| (|01\kt -|10\kt)=(-1)^i \delta_{i_1,i_2 \oplus 1}$, where the $\oplus$
denotes an addition modulo 2.
\begin{equation}
\br i_{2k+1} i_{2k+2} \cdots i_N |R_N \kt=\dfrac{(-1)^{i_{2k+1}+i_{2k+3}+\cdots+i_{N-1}}}{(\sqrt{2})^{(N-2k)/2}} \delta_{i_{2k+1},i_{2k+2} \oplus 1} \cdots \delta_{i_{N-1},i_{N}\oplus 1} |R_{2k}\kt 
\label{rhocross1}
\end{equation}
and a similar expression is found for $\br L_N | i_{2k+1} i_{2k+2} \cdots i_N \rangle$ which involves the untraced part $|L_{2k-2}\kt$ as follows,
\begin{eqnarray}
\br L_N|i_{2k+1} i_{2k+2} \cdots i_N \kt&=&\br L_{2k-2}| \bigg(\frac{1}{\sqrt{2}}\bigg)^{((N-2k)/2)-2} \delta_{i_{2k+2} i_{2k+3}\oplus1} \cdots \delta_{i_{N-2} i_{N-1}\oplus1}  \times \nonumber \\
&&\br (2k~2k+1) |i_{2k+1}\kt \br (N 1) |i_{N}\kt  %% i have used ampersand inside ampersand method to break - useful technique %%
\label{rhocross2}
\end{eqnarray}
The ``end spins" are taken into account as $\br (2k~2k+1) |i_{2k+1}\kt=\frac{1}{\sqrt{2}}(\br 0|_{2k} \delta_{1,i_{2k+1}}-\br 1|_{2k} \delta_{0,i_{2k+1}})$ and $\br (N 1) |i_{N}\kt=\frac{1}{\sqrt{2}}(\br 1|_{1} \delta_{0,i_N}-\br 0|_{1} \delta_{1,i_N})$.
Using this expression along with Eqs.~(\ref{rhocross1},\ref{rhocross2}) and substituting them in Eq.~(\ref{rhocross}), the final form of $\rho_{2k}$ (after some straightforward algebra taking care of the modulo 2 addition) is found to be
\beq
\begin{split}
\rho_{2k}& =\alpha_1^2 \vert R_{2k} \kt \br R_{2k} \vert+\alpha_2^2 \bigg[ \dfrac{I_{1}}{2} \otimes |L_{2k-2}\kt\br L_{2k-2}| \otimes \dfrac{I_{2k}}{2} \bigg]  \\
& + \dfrac{\alpha_1 \alpha_2}{2^{(N-2k)/2}}(-1)^{(N-2k)/2}\bigg[\vert R_{2k} \kt \br L_{2k}| + \vert L_{2k} \kt \br R_{2k}|\bigg].
\end{split}
\label{dimerRDMfinal}
\eeq
It is to be noted from the structure of $\rho_{2k}$ that the coherent term is only of the order of  $2^{-(N-2k)/2}$ and hence exponentially decreases with the number of spins in block $B$.

The eigenvalues of $\rho_{2k}$ are now calculated. For notational convenience, for $K$ spins, define $p=\br R_K | L_K\kt=(-1)^{K/2} 2^{(1-K/2)}$ and $\gamma=\dfrac{\alpha_1 \alpha_2 (-1)^{(N-2k)/2}}{2^{(N-2k)/2}}$. Writing the $I_1\otimes I_{2k}$ in $\bigg(I_1 \otimes |L_{2k-2}\kt \br L_{2k-2}| \otimes I_{2k}\bigg)$ as the sum of projectors into the four Bell basis, $|\phi_b^l\kt$ ($1\le l \le 4$)  we can define new states $|L_{2k}^l\kt =|L_{2k-2}\kt \otimes |\phi_b^l\kt _{2k,1}$, here $l=1,2,3,4$, we then use the (easily obtained) properties that $\br R_{2k}|L_{2k}^l \kt=\br L_{2k}|L_{2k}^l \kt=0, \; l=1,2,3$. Explicitly: $|L_{2k}^1\kt=|L_{2k-2}\kt \big((|01\kt+|10\kt)/\sqrt{2}\big)_{2k,1}$~,~$|L_{2k}^2\kt=|L_{2k-2}\kt \big((|00\kt+|11\kt)/\sqrt{2}\big)_{2k,1}$~,~$|L_{2k}^3\kt=|L_{2k-2}\kt \big((|00\kt-|11\kt)/\sqrt{2}\big)_{2k,1}$ and $|L_{2k}^4\kt=|L_{2k-2}\kt \big((|01\kt-10\kt)/\sqrt{2}\big)_{2k,1}=|L_{2k}\kt$. Now using the above we can rewrite $\rho_{2k}$ as
\begin{equation}
\rho_{2k}=\alpha_1^2 |R_{2k} \kt \br R_{2k}| +\dfrac{\alpha_2^2}{4}|L_{2k} \kt \br L_{2k}| + \dfrac{\alpha_2^2}{4}\bigg[\sum_{l=1}^3 |L_{2k}^l \kt \br L_{2k}^l|\bigg]+
\gamma \bigg[|R_{2k} \kt \br L_{2k} |+|L_{2k} \kt \br R_{2k} |\bigg] 
\label{bellrdm}
\end{equation}
It is straightforward to verify that $|L_{2k}^1\kt$,$|L_{2k}^2\kt$ and $|L_{2k}^3\kt$ are three eigenstates of $\rho_{2k}$ with degenerate eigenvalues  $\dfrac{\alpha_2^2}{4}$. This corresponds to the triply degenerate eigenvalue in the entanglement spectrum at the MG point. 
It is clear that the other eigenvalues correspond to eigenvectors in the two dimensional subspace spanned by $|R_{2k}\kt$ and $|L_{2k}\kt$,
and hence there are only two of these. Either defining the orthogonal vectors $|R_{2k}\kt \pm |L_{2k}\kt$ or proceeding to define linear
superpositions of  $|R_{2k}\kt$ and $|L_{2k}\kt$ as the eigenvectors, a straightforward (if somewhat lengthy) calculation leads to Eq.~(\ref{dimerevals12}). These two eigenvalues correspond to the most dominant state and the rising state in the entanglement spectrum of $\rho_{2k}$.

%\newpage \bibliographystyle{rkbib}

\bibliography{myref.bib}

\end{document}